\documentclass[
reprint,
superscriptaddress,
showpacs,showkeys,preprintnumbers,
amsmath,amssymb,
aps,
prl,
]{revtex4-1}

\usepackage{graphicx}% Include figure files
\usepackage{dcolumn}% Align table columns on decimal point
\usepackage{bm}% bold math
\usepackage{hyperref}% add hypertext capabilities
\usepackage{enumerate}

\usepackage{titlesec}

\titlespacing\section{0pt}{12pt plus 4pt minus 2pt}{0pt plus 0pt minus 0pt}
\titlespacing\subsection{0pt}{12pt plus 4pt minus 2pt}{0pt plus 2pt minus 2pt}
\titlespacing\subsubsection{0pt}{12pt plus 4pt minus 2pt}{0pt plus 2pt minus 2pt}
\usepackage{dsfont}
\DeclareMathOperator{\Tr}{Tr}
\begin{document}

%\preprint{APS/123-QED}

\title{Entanglement and Designs}

\author{Matthew A.\ Graydon}
\email{mgraydon[at]perimeterinstitute.ca}
\affiliation{Perimeter Institute for Theoretical Physics, 31 Caroline St.\ N., Waterloo, Ontario N2L 2Y5, Canada}
\affiliation{Institute for Quantum Computing, University of Waterloo, Waterloo, Ontario N2L 3G1, Canada}
\affiliation{Department of Physics $\&$ Astronomy, University of Waterloo, Waterloo, Ontario N2L 3G1, Canada}
\author{D.\ M.\ Appleby}
\email{marcus[at]physics.usyd.edu}
\affiliation{Centre for Engineered Quantum Systems, School of Physics, The University of Sydney, Sydney, NSW 2006, Australia}
\date{\today}
\begin{abstract}
We describe a connection between entanglement and designs.   It involves the conical 2-designs  introduced in a previous  paper.  These are a generalization of projective 2-designs which includes full sets of arbitrary rank mutually unbiased measurements (\textsc{mum}s) and arbitrary rank symmetric informationally complete measurements (\textsc{sim}s), as well as the more familiar \textsc{mub}s and \textsc{sic}s. We show that a \textsc{povm} is a conical 2-design if and only if there exists what we call a regular entanglement monotone whose restriction to the pure states is a function of the norm of the probability vector.  In that case the concurrence is such a monotone.  We also generalize and develop previous work on designs and entanglement detection.
\end{abstract}

% Classification Scheme.

\maketitle
\phantomsection
\section{I. INTRODUCTION}
Entanglement is fundamental in quantum physics.  Complex projective $2$-designs---geometric structures such as full sets of \textsc{mub}s (mutually unbiased bases)~\cite{Schwinger:1960,Ivanovic81,Wootters89,Durt:2010} and \textsc{sic}s (symmetric informationally complete measurements of unit rank)~\cite{Zauner99,Renes04A,Scott10}---are also important. Recently some connections have been found between the two areas ~\cite{Wiesniak:2011,Spengler12,Chen:2014,Chen:2015,Liu:2015,Shen:2015,Kalev:2013}. We describe a quite  different connection.

The connection involves the concurrence \cite{Wootters98b,Rungta01,Rungta03}.  This is an entanglement monotone \cite{Vidal00} which is simply related to the entanglement of formation \cite{Bennett:1996} and which has some useful properties. Computing entanglement monotones for an arbitrary mixed state can be a difficult problem.  The concurrence has the useful feature that it has easily computable lower and upper bounds~\cite{Mintert04,Minert:2004}.  In the case of a pair of qubits there is an explicit formula for the concurrence itself~\cite{Wootters98b}.  Another useful property is the fact that it vanishes if and only if the state is separable. Finally, it illuminates the monogamous character of entanglement~\cite{Coffman00}.

The connection also involves conical designs~\cite{Graydon:2015}.  A complex projective $2$-design \cite{Neumaier81, Hoggar82, Zauner99, Scott06} is a set of pure states $\Pi_1,\dots, \Pi_m$ such that $\sum_{\alpha} \Pi_{\alpha} \otimes \Pi_{\alpha}$ is proportional to the projector onto the symmetric subspace of $\mathcal{H}\otimes \mathcal{H}$ (where $\mathcal{H}$ is a fixed $d$ dimensional complex Hilbert space).  A conical $2$-design is a generalization in which the rank 1 and trace 1 conditions are relaxed.  Specifically it is a set of positive semi-definite operators $E_1, \dots, E_m$ such that
\begin{align}
\sum_{\alpha}E_{\alpha}\otimes E_{\alpha}&=k_{s}\Pi_{\text{sym}}+k_{a}\Pi_{\text{asym}}, & k_{s}&>k_{a}\geq 0
\label{qcdDef}
\end{align}  
where $\Pi_{\text{sym}}$ and $\Pi_{\text{asym}}$ are respectively the projectors onto the symmetric and anti-symmetric subspaces of $\mathcal{H}\otimes \mathcal{H}$.  It can be shown that $k_a =0$ if and only if $E_{\alpha}$ are all rank 1, in which case they form a weighted complex projective 2-design (up to scaling).  Full sets of arbitrary rank mutually unbiased measurements (\textsc{mum}s)~\cite{Kalev14b} and arbitrary rank symmetric informationally complete measurements (\textsc{sim}s)~\cite{Appleby07} are conical 2-designs.

We will need the following topological concept. Let  $\mathcal{S}$ be the unit sphere in $\mathcal{H}\otimes \mathcal{H}$.  We  say that an entanglement monotone $M$ is regular if for all $x$ the set $ \{|\Psi\rangle \in \mathcal{S} \colon M(|\Psi\rangle)=x\}$ has  empty interior (i.e.\ every point in the set is the limit of a sequence of points outside it~\cite{Dugundji:1966}).  Observe that if $\{|\Psi\rangle \in \mathcal{S} \colon M(|\Psi\rangle)=x\}$ has $\mu$-measure zero for all $x$, then the monotone is regular ($\mu$ being the usual invariant measure on $\mathcal{S}$).   The condition is satisfied by a wide variety of monotones.  In particular, the concurrence is regular.

We are now ready to state our main result.  Let $E_{\alpha}$ be an arbitrary \textsc{povm} on $\mathcal{H}$.  Let $|\Psi\rangle\in \mathcal{S}$, 
\begin{equation}
p_{\alpha,\beta} = \langle \Psi | E_{\alpha} \otimes E_{\beta} |\Psi\rangle
\end{equation}
be the outcome probabilities for the bipartite product measurement $E_{\alpha}\otimes E_{\beta}$,
\begin{equation}
\| \vec{p} \|=\sqrt{\sum_{\alpha,\beta}p_{\alpha,\beta}^{2}}
\label{euclid}
\end{equation}
be the norm of the probability vector, and $C(|\Psi\rangle)$ be the concurrence.

\vspace{6pt}
\noindent\label{Theorem1}\textbf{Theorem 1.}  $E_{\alpha}$ is a conical $2$-design if and only if there exists a regular entanglement monotone whose restriction to $\mathcal{S}$ is a function of $\|\vec{p}\|$.  In that case the restriction of the concurrence  is a function of $\|\vec{p}\|$.  Specifically
\begin{equation}
C(|\Psi\rangle) = 2\sqrt{\frac{\hspace{0.34cm}k_s^2- \|\vec{p}\|^2}{k_s^2-k_a^2}} \text{.}
\label{conSimplified}
\end{equation}

\vspace{6pt}
Given that $C(|\Psi\rangle)^2$ is necessarily quadratic in the probabilities it seems fair to say that Eq.~(\ref{conSimplified}) is about the simplest expression conceivable.  A simple description of entanglement in terms of probabilities is important in those theoretical approaches which seek to formulate quantum mechanics in purely probabilistic language~\cite{Fuchs10,Fuchs13,Barnum12}.  The simplicity of the result---the fact that conical designs are naturally adapted to the description of  entanglement---means that  it is likely to be important for other reasons also.

It may be possible to strengthen theorem~\hyperref[Theorem1]{1}, so that it states that $E_{\alpha}$ is a conical 2-design if and only if there is any monotone at all whose restriction to $\mathcal{S}$ is a function of $\|\vec{p}\|$.  However, we have not been able to prove it.

The next three sections are devoted to the proof of theorem~\hyperref[Theorem1]{1}.  In Section~\hyperref[secV]{V} we generalize the results of refs.~\cite{Spengler12,Chen:2014,Chen:2015} to the case of an arbitrary conical $2$-design, and compare them to the result just described.

\phantomsection
\label{secII}\section[II. THEOREM 1:  NECESSITY]{II. THEOREM~\hyperref[Theorem1]{1}:  NECESSITY}Let $E_{\alpha}$ be a \textsc{povm} which is also a conical $2$-design, and let $|\Psi\rangle\in \mathcal{S}$ have Schmidt decomposition
\begin{equation}
|\Psi\rangle=\sum_{r}  \lambda_{r} \big(|f_{r}\rangle\otimes|g_{r}\rangle\big), \quad \sum_r \lambda^2_r =1, \quad \lambda_r \ge 0
\label{schmidt}
\end{equation}  
where $|f_{r}\rangle$, $|g_{r}\rangle$ are \textsc{onb}s. Then Eq.~\eqref{qcdDef} implies
\begin{align}
\|\vec{p}\|^{2}&=\sum_{\alpha,\beta}\langle\Psi|E_{\alpha}\otimes E_{\beta}|\Psi\rangle^{2}\nonumber
\\
&=\langle\Psi|\otimes \langle \Psi| \sum_{\alpha,\beta} (E_{\alpha}\otimes E_{\beta})\otimes (E_{\alpha}\otimes E_{\beta})|\Psi\rangle \otimes |\Psi\rangle \nonumber
\\
&=\sum_{r,s,u,v}\lambda_{r}\lambda_{s}\lambda_{u}\lambda_{v}\left(\frac{k_{+}\delta_{rs}\delta_{uv}}{2}+\frac{k_{-}\delta_{ru}\delta_{sv}}{2}\right)^{2}\nonumber\\
&=\frac{k_{s}^{2}+k_{a}^{2}}{2}+\frac{k_{s}^{2}-k_{a}^{2}}{2}\sum_{r=1}^{d}\lambda_{r}^{4}
\label{pNorm}
\end{align}
where $k_{\pm} = k_s \pm k_a$. Consequently~\cite{Rungta01} (with $\rho$ being the reduced density matrix of either of the two subsystems)
\begin{equation}
C(|\Psi\rangle) = \sqrt{2-2\mathrm{Tr}\big(\rho^{2}\big)}=2\sqrt{\frac{\hspace{0.34cm}k_s^2-\|\vec{p}\|^2}{k_s^2-k_a^2} } 
\end{equation}

\phantomsection
\label{secIII}\section{III. LOCAL UNITARY INVARIANCE}
In this section we prove a subsidiary result which is needed to complete the proof of theorem~\hyperref[Theorem1]{1}. Define a  non-trivial local unitary invariant to be a function $f$ on $\mathcal{S}$ which is (a) not constant and (b) such that $f(U\otimes V |\Psi\rangle) = f(|\Psi\rangle)$ for all $|\Psi\rangle$ and $U, V\in \text{U}(d)$ (where $\text{U}(d)$ is the unitary group in dimension $d$). 

\vspace{6pt}
\noindent\label{Lemma2}\textbf{Lemma 2.} $E_{\alpha}$ is a conical 2-design if and only if $\|\vec{p}\|$ is a non-trivial local unitary invariant.  

\vspace{6pt}
\noindent\emph{Proof.} Necessity is an immediate consequence of Eq.~\eqref{pNorm} together with the fact that $k_s > k_a$.
To prove sufficiency, let $E_{\alpha}$ be a \textsc{povm} such that $\|\vec{p}\|$ is a non-trivial local unitary invariant.  Define, for all $U\in \text{U}(d)$,
\begin{align}
E^U_{\alpha} &=U^{\dagger} E_{\alpha} U \text{,} & N^U &= \sum_{\alpha} E^U_{\alpha}\otimes E^U_{\alpha}\text{.}
\end{align}
We will show that $N^U$ is independent of $U$.  

Let $W_{23}$ be the unitary operator which swaps the second and third factors in $\mathcal{H}\otimes \mathcal{H} \otimes \mathcal{H}\otimes \mathcal{H}$.  Local unitary invariance means
\begin{align}
\|\vec{p}\|^2 &= \langle \Psi | \otimes \langle \Psi | \; W_{23} (N^U\otimes N^V) W_{23} \; | \Psi\rangle \otimes |\Psi\rangle 
\label{pvcTermsNUNV}
\end{align}
is independent of $U,V$ for all $|\Psi\rangle$.  Note that this remains true even if $|\Psi\rangle$ is not normalized (although $\vec{p}$ then loses its interpretation as a probability vector).

Let $|\psi_0\rangle$, $|\psi_1\rangle$ be arbitrary vectors in $\mathcal{H}$ and  $e^{i\theta}$  an arbitrary phase.   Define
\begin{equation}
 |\Psi\rangle = \sum_{n=0}^1 e^{ni\theta} |\psi_n\rangle \otimes |\psi_n \rangle \text{.}
\end{equation}
Then
\begin{equation}
W_{23}\Bigl(|\Psi\rangle \otimes |\Psi\rangle\Bigr) = \sum_{n=0}^2 e^{in\theta}|\Xi_n\rangle
\end{equation}
where 
we have defined $|\Xi_{0}\rangle=|\psi_{0}\rangle\otimes|\psi_{0}\rangle\otimes|\psi_{0}\rangle\otimes|\psi_{0}\rangle$, $|\Xi_{1}\rangle=|\psi_{0}\rangle\otimes|\psi_{1}\rangle\otimes|\psi_{0}\rangle\otimes|\psi_{1}\rangle +|\psi_{1}\rangle\otimes|\psi_{0}\rangle\otimes|\psi_{1}\rangle\otimes|\psi_{0}\rangle$, and $|\Xi_{2}\rangle=|\psi_{1}\rangle\otimes|\psi_{1}\rangle\otimes|\psi_{1}\rangle\otimes|\psi_{1}\rangle$.
Substituting into Eq.~\eqref{pvcTermsNUNV} gives
\begin{equation}
\|\vec{p}\|^2 = \sum_{n=-2}^2 A_n(U,V) e^{in \theta}
\end{equation}
where
\begin{align}
A_n(U,V)=  \sum_{r,s=0}^2 \delta_{s,r+n} \langle \Xi_r|N^U\otimes N^V| \Xi_s\rangle \text{.}
\end{align}
The arbitrariness of $e^{i\theta}$ means that the coefficients in this expansion are independent of $U,V$.  In particular
\begin{equation}
A_0(U,V) = \sum_{n=0}^2 \langle \Xi_n|  N^U \otimes N^V |\Xi_n\rangle
\label{A0Expn}
\end{equation}
is independent of $U$, $V$. Setting $|\psi_1\rangle=0$ (respectively, $|\psi_0\rangle=0$) in Eq.~\eqref{A0Expn} shows that the $n=0$ (respectively, $n=2$) term on the right hand side is independent of $U$, $V$.  So  all three terms are independent of $U$, $V$.  Now observe that $N^U$ commutes with $W_{12}$, the unitary operator which swaps the two factors in $\mathcal{H}\otimes \mathcal{H}$.  It follows that, if we define 
\begin{equation}
\vec{L}^U = \begin{pmatrix} \langle \psi_0|\otimes \langle\psi_1 | N^U | \psi_0\rangle\otimes |\psi_1\rangle \\  \langle\psi_0|\otimes\langle \psi_1 | N^U | \psi_1\rangle\otimes |\psi_0\rangle \end{pmatrix} \text{,}
\end{equation}
then $\vec{L}^U \cdot \vec{L}^V$ is independent of $U$, $V$.   So
\begin{equation}
\vec{L}^U \cdot \vec{L}^U  = \vec{L}^V \cdot \vec{L}^V = \vec{L}^U \cdot \vec{L}^V \text{.}
\end{equation}
for all $U$, $V$.  Since the components of $\vec{L}^U$, $\vec{L}^V$ are real, non-negative this means $\vec{L}^U=\vec{L}^V$.   In particular
\begin{equation}
\langle \psi_0| \otimes \langle\psi_1  | N^U |\psi_0\rangle \otimes |\psi_1 \rangle
\end{equation}
is independent of $U$.

Now let $|\phi_0\rangle, |\phi_1\rangle, |\chi_1\rangle, |\chi_2\rangle$ be arbitrary vectors in $\mathcal{H}$ and  $e^{i\theta}$  an arbitrary phase.  Define
\begin{align}
|\psi_0\rangle &= \sum_{n=0}^1 e^{in\theta}| \phi_n \rangle & |\psi_1\rangle &= \sum_{n=0}^1 e^{2in\theta} |\chi_n\rangle
\end{align}
Then
\begin{equation}
\langle \psi_0| \otimes \langle\psi_1  | N^U |\psi_0\rangle \otimes |\psi_1 \rangle = \sum_{n=-3}^3 B_n(U) e^{in \theta}
\end{equation}
where
\begin{align}
&B_n(U) 
&= \sum_{\substack{r_2+2s_2=\\ r_1+2s_1+n}} \langle \phi_{r_1}|\otimes \langle\chi_{s_1} | N^U | \phi_{r_2}\rangle \otimes |\chi_{s_2}\rangle
\end{align}
The arbitrariness of $e^{i\theta}$ means that the individual coefficients in this expansion must be independent of $U$.  In particular
\begin{equation}
B_3(U) = \langle \phi_0| \otimes \langle\chi_0 | N^U | \phi_1\rangle \otimes |\chi_1\rangle
\end{equation}
is independent of $U$.  Since the product states span $\mathcal{H}\otimes \mathcal{H}$ this means $N^U$ is independent of $U$.  In other words $\sum_{\alpha} E_{\alpha} \otimes E_{\alpha}$ commutes with $U\otimes U$ for all $U$.  Theorem~1 in ref.~\cite{Graydon:2015} then implies
\begin{equation}
\sum_{\alpha} E_{\alpha}\otimes E_{\alpha} = k_s \Pi_{\text{sym}} + k_a \Pi_{\text{asym}}
\end{equation}
for some $k_s \ge k_a \ge 0$. If $k_s=k_a$ then $\sum_{\alpha} E_{\alpha} \otimes E_{\alpha} = k_s I$, implying $\|\vec{p}\| = k_s$ for all $\|\vec{p}\|$, contradicting the fact that $\| \vec{p}\|$ is a non-trivial invariant.  So $k_s> k_a$ and $E_{\alpha}$ is a conical 2-design.  $\square$

\phantomsection
\label{secIV}\section[IV. THEOREM 1:  SUFFICIENCY]{IV. THEOREM~\hyperref[Theorem1]{1}:  SUFFICIENCY}
Suppose $M$ is a regular entanglement monotone such that  $M=f(\|\vec{p}\|)$ on $\mathcal{S}$, for some function $f$.  
We first show that $\|\vec{p}\|$ must be a non-trivial local unitary invariant.  For suppose that were not the case.  Then either $\|\vec{p}\|$ would be trivial or else it would  not be a local unitary invariant.  The first possibility  would imply that $M$ was constant on $\mathcal{S}$, contradicting the fact that $M$ is assumed regular.  The second possibility would imply that, for some fixed $|\Psi\rangle$, the map
\begin{align}
&(U,V) \in \text{U}(d)\times\text{U}(d) 
\nonumber
\\
& \hspace{0.1 cm} \to \sqrt{\langle \Psi | \otimes \langle \Psi | \; W_{23} (N^U\otimes N^V) W_{23} \; | \Psi\rangle \otimes |\Psi \rangle}\in \mathbb{R} 
\end{align}
took at least two distinct values, say $a< b$.  The continuity of the map together with the connectedness of the group $\text{U}(d)\times \text{U}(d)$ would then imply that it took every value in the interval $[a,b]$.  On the other hand, the restriction of $M$ to $\mathcal{S}$ is a local unitary invariant~\cite{Vidal00}.  So $f$ would have to take the same constant value, call it $x$, for every element of $[a,b]$.  This would mean that $\{|\Psi\rangle \in \mathcal{S} \colon M(|\Psi\rangle)=x\})$ contained the non-empty open set
$\{|\Psi\rangle \in \mathcal{S}: a < \| \vec{p} \| < b \}$, again contradicting the fact that $M$ is assumed regular.  
It now follows from Lemma~\hyperref[Lemma2]{2}  that $E_{\alpha}$ is a conical 2-design.

\phantomsection
\label{secV}\section{V. WITNESSES}
Having established our main result we now compare it with the results in refs.~\cite{Spengler12,Chen:2014,Chen:2015,Liu:2015,Shen:2015}.  We begin with refs.~\cite{Spengler12,Chen:2014,Chen:2015,Liu:2015}. 
Although the authors do not present it  this way what is done in these papers is, in effect,  to show that there is a natural way to construct entanglement witnesses out of \textsc{mum}s and \textsc{sim}s (as noted in ref.~\cite{Chruscinski14}).
Let $\mathcal{S}_s$ be the subset of $\mathcal{S}$ consisting of the separable pure states.  Given an arbitrary Hermitian operator $A$ on $\mathcal{H}\otimes \mathcal{H}$ define
\begin{align}
s_A^{-} &= \inf_{|\Psi\rangle \in \mathcal{S}_s} \bigl\{\langle \Psi|A|\Psi\rangle\bigr\} & s_A^{+} &= \sup_{|\Psi\rangle \in \mathcal{S}_s} \bigl\{\langle \Psi|A|\Psi\rangle\bigr\}
\\
e_A^{-} &= \inf_{|\Psi\rangle \in \mathcal{S}} \bigl\{\langle \Psi|A|\Psi\rangle\bigr\} & e_A^{+} &= \sup_{|\Psi\rangle \in \mathcal{S}} \bigl\{\langle \Psi|A|\Psi\rangle\bigr\}
\end{align}
Then $s_A^{+}I - A$ (respectively $A-s_A^{-} I$) is an entanglement witness~\cite{Chruscinski14,Guhne:2009} if and only if $e_A^{+}> s_A^{+}$ (respectively $e_A^{-}< s_A^{-}$), in which case we will say $A$ detects entanglement from above (respectively below).  If, on the other hand, $s_A^{\pm}= e_A^{\pm}$, then a measurement of $A$ cannot detect entanglement. In refs.~\cite{Spengler12,Chen:2014,Chen:2015} the authors only consider detection from above, although detection from below is also possible, as we will see.

Let $E_{\alpha}$ be an arbitrary conical $2$-design and let $|e_j\rangle$ be some fixed \textsc{onb} in $\mathcal{H}$.  Define
\begin{align}
N &= \sum_{\alpha}E_{\alpha} \otimes E_{\alpha}\text{,} & N^{\rm{PT}} &= \sum_{\alpha} E_{\alpha} \otimes E^{\rm{T}}_{\alpha}\text{,}
\end{align}
where $N^{\rm{PT}}$ (respectively $E^{\rm{T}}_{\alpha}$) is the partial transpose (respectively transpose) of $N$ (respectively $E_{\alpha}$) relative to the basis $|e_j\rangle$.  It follows from Eq.~\eqref{qcdDef} in this paper and Theorem~1 in ref.~\cite{Graydon:2015} that
\begin{align}
N& = k_{+} I + k_{-} W_{12}\text{,} &  N^{\rm{PT}} &= k_{+}I + d k_{-} |\Phi_{+}\rangle \langle \Phi_{+} | \text{,}
\end{align}
where $k_{\pm} = (k_s \pm k_a)/2$, $W_{12}$ is the unitary which swaps the two factors in $\mathcal{H}\otimes \mathcal{H}$, and $|\Phi_{+}\rangle$ is the maximally entangled state $(1/\sqrt{d})\sum_{j} |e_j\rangle \otimes |e_j\rangle$.  One easily sees
\begin{align}
s_N^{-}  &= k_{+}\text{,}  & s_N^{+} &= k_{+} + k_{-}\text{,}
\\
e_N^{-}  &= k_{+}-k_{-}\text{,} &  e_N^{+} &= k_{+}+k_{-}\text{.}
\end{align}
So $N$ can detect entanglement from below, but not from above.  On the other hand
\begin{align}
s_{N^{\rm{PT}}}^{-}  &=k_{+} \text{,} & s_{N^{\rm{PT}}}^{+}  &= k_{+}+k_{-}\text{,}
\\
e_{N^{\rm{PT}}}^{-} &= k_{+}\text{,} & e_{N^{\rm{PT}}}^{+}  &= k_{+}+dk_{-}\text{.}
\end{align}
So $N^{\rm{PT}}$ can detect entanglement from above, but not from below.  
Let us note that in refs.~\cite{Spengler12,Chen:2014,Chen:2015,Liu:2015} the authors calculate $c^s_{+}(N)$ for \textsc{mum}s and \textsc{sim}s, but not $c_{+}(N)$.  Consequently, they do not draw the conclusion, that $N$ cannot detect entanglement from above.

In refs.~\cite{Spengler12,Chen:2014,Chen:2015,Liu:2015} the authors also consider operators of the more general form
\begin{align}
N_{\rm{gen}} &= \sum_{\alpha} E_{\alpha}\otimes F_{\alpha}
\end{align}
where $E_{\alpha}$, $F_{\alpha}$ are distinct \textsc{mum}s or \textsc{sim}s.  Calculating  $c_{\pm}^s(N_{\rm{gen}})$, $c_{\pm}(N_{\rm{gen}})$ for arbitrary  pairs of conical designs is beyond our present scope.  In this connection let us note that the authors of refs.~\cite{Spengler12,Chen:2014,Chen:2015,Liu:2015} do not calculate them either (although refs.~\cite{Spengler12,Chen:2014,Chen:2015} do calculate a bound for $c_{+}^s(N_{\rm{gen}})$ valid for pairs of \textsc{mum}s or \textsc{sim}s having the same  contraction parameter~\cite{Graydon:2015}, extended in ref.~\cite{Liu:2015}   to pairs of \textsc{mum}s  having different contraction parameters).  

Liu \emph{et al}~\cite{Liu:2015}  consider the application of \textsc{mum}s to multipartite entanglement.  However, they do not show that their  bound is violated for any non-separable states.

Shen \emph{et al}~\cite{Shen:2015} consider detection  criteria which are non-linear in the density matrix.  It is easily seen that their criteria generalize to the statements that, for any conical design, and any separable density matrix $\rho$, the quantities $\Bigl| \Tr\bigl(N(\rho-\rho_1\otimes \rho_2)\bigr) \Bigr|$, $\Bigl| \Tr\bigl(N^{\rm{PT}}(\rho-\rho_1\otimes \rho_2)\bigr) \Bigr|$ are both bounded above by $k_{-} \sqrt{\bigl(1-\Tr(\rho_1^2)\bigr)\bigl(1-\Tr(\rho_2^2)\bigr)}$,
where $\rho_j$ is the reduced density matrix for the $j^{\rm{th}}$ subsystem.  It is easily verified that every entangled state detected by the criterion $\Tr(\rho N^{\rm{PT}})> s_{N^{\rm{PT}}}^{+}$ is also detected by the corresponding quadratic criterion.  On the other hand there exist states which are detected by the  criterion $\Tr(\rho N) < s_N^{-}$  but which are not detected by the corresponding quadratic criterion.   Consider, for instance, the entangled Werner state~\cite{Werner:1989}
\begin{align}
\rho_{W} &= \frac{2(1-p)}{d(d+1)}\Pi_{\text{sym}} + \frac{2p}{d(d-1)} \Pi_{\text{asym}} & \frac{1}{2} < p \le 1
\end{align}
The linear criterion involving $N$ detects the entanglement for all values of $p$ whereas  the quadratic criterion only detects it for $p> (d-1)/d$.  This does not conflict with Shen \emph{et al}'s analysis because they do not consider the possibility of detection from below.

The witnesses corresponding to $N$ and $N^{\rm{PT}}$ are
\begin{align}
N-s_N^{-} I &= k_{-} W_{12}  
\\
s_{N^{\rm{PT}}}^{+} I - N^{\rm{PT}} &= k_{-}\bigl( I -d |\Phi_{+}\rangle \langle \Phi_{+} |\bigr)
\end{align}
The fact that these are witnesses is, of course, well known (see, e.g., Example 3.1 in ref.~\cite{Chruscinski14};  Eq.~(28) in ref.~\cite{Guhne:2009}).  The novelty of refs.~\cite{Spengler12,Chen:2014,Chen:2015,Liu:2015} is that they show that the witnesses have  simple expressions in terms of the probabilities obtained from local measurements.  The simplicity of the expressions means they have an obvious theoretical interest.  Their experimental interest is less obvious, since obtaining the probabilities empirically amounts to performing full-state tomography; and once one has done that one can calculate any witness one wants.  It is to be observed, however, that one may be able to truncate the design and still have an entanglement witness (as is shown  in Appendix A of Spengler \emph{et al}~\cite{Spengler12} for  \textsc{mub}s).  As Spengler \emph{et al} note this may be experimentally useful, especially when $d$ is large.

To summarize, we have shown that a well-known entanglement monotone has a simple expression in terms of design probabilities while refs.~\cite{Spengler12,Chen:2014,Chen:2015} have done the same for two well-known entanglement witnesses.  The crucial difference is that theorem~\hyperref[Theorem1]{1} gives a condition which is both necessary and sufficient for a given \textsc{povm} to be a conical 2-design, whereas continuity means that the inequalities $s_N^{-} > e_N^{-} $, $s_{N^{\rm{PT}}}^{+} < e_{N^{\rm{PT}}}^{+}$ will remain valid even after the $E_{\alpha}$ have been significantly and randomly perturbed.  Consequently, the inequalities proved in refs.~\cite{Spengler12,Chen:2014,Chen:2015,Liu:2015,Shen:2015} can detect entanglement for a wide variety of \textsc{povm}s which are not conical 2-designs.   In that sense the connection we exhibit, between designs and entanglement, is tighter than the one exhibited in refs.~\cite{Spengler12,Chen:2014,Chen:2015,Liu:2015,Shen:2015}.

\phantomsection
\section{CONCLUSION}
We showed that a \textsc{povm} is a conical 2-design if and only if there is a regular entanglement monotone which is a function of $\|\vec{p}\|$.  We went on to extend the results in refs.~\cite{Spengler12,Chen:2014,Chen:2015,Liu:2015,Shen:2015}, and to compare them with our theorem~\hyperref[Theorem1]{1}.  In particular we showed that there is a natural way to construct entanglement witnesses out of an arbitrary conical design.  However, the connection between witnesses and designs is less tight than the one between monotones and designs, in the sense explained in the last section.  Our work naturally suggests the question, whether there are similar connections between multipartite entanglement and conical $t$-designs with $t>2$.

\phantomsection
\section{ACKNOWLEDGEMENTS}
We gratefully acknowledge Chris Fuchs for stimulating discussions. Research at Perimeter Institute is supported by the Government of Canada through Industry Canada and by the Province of Ontario through the Ministry of Research \& \mbox{Innovation}. MAG was supported by an NSERC Alexander Graham Bell Canada Graduate Scholarship. DMA was supported by the IARPA MQCO program, by the ARC via EQuS project number CE11001013, and by the US Army Research Office grant numbers W911NF-14-1-0098 and W911NF-14-1-0103.
\bibliography{cdRefV2}
\end{document}